\documentclass[twocolumn, journal]{IEEEtran}
 
\usepackage{booktabs} 
\usepackage{multirow}
\usepackage{mathtools}

\usepackage{circuitikz}

\usepackage{nccmath}

\usepackage{amssymb}
 
\usepackage{enumitem}

\usepackage{smartdiagram}
\usesmartdiagramlibrary{additions}

\graphicspath{{Figs/}}

\usepackage{pdfpages}
\usepackage{pdflscape}
\usepackage{amsmath}
\usepackage{amsthm}

\usepackage{amsfonts}

\usepackage{color}

\usepackage{setspace}
\usepackage{enumitem}
\usepackage{xcolor}

\usepackage{bm}
\usepackage{bbm}

\usepackage{tikz}
\usepackage{pgf}
\usetikzlibrary{arrows,automata}
\usetikzlibrary{shapes}
\tikzstyle{data44}=[rectangle split,rectangle split parts=2,draw,text centered]

\usepackage{subfigure}

\DeclareMathAlphabet{\mathcal}{OMS}{cmsy}{m}{n}

\usetikzlibrary{matrix}

\tikzset{
  BarreStyle/.style =   {opacity=.3,line width=14 mm,color=#1},
  node style ge/.style={},
  node style sp/.style={},
  yl/.style={},
  arrow style mul/.style={},
}

\newtheoremstyle{mytheoremstyle} 
        {\topsep}                    
        {\topsep}                    
        {\fontfamily{ptm}\selectfont}                   
        {}                           
        {\itshape\fontfamily{ptm}\selectfont}                   
        {:}                          
        {.5em}                       
        {}  
\theoremstyle{mytheoremstyle}

\usepackage{threeparttable,booktabs}

\usepackage[colorlinks,
            linkcolor=blue,
            anchorcolor=blue,
            citecolor=blue
            ]{hyperref}

\usepackage{cite}

\usepackage{mdwlist}

\usepackage[linesnumbered,ruled,vlined]{algorithm2e}
\SetKwInput{KwInput}{Input}                
\SetKwInput{KwOutput}{Output}   
\let\oldnl\nl
\newcommand{\nonl}{\renewcommand{\nl}{\let\nl\oldnl}}

\usepackage{etoolbox}

\makeatletter
\patchcmd{\@algocf@start}
  {-1.5em}
  {0pt}
  {}{}
\makeatother

\newcommand{\hig}[1]{\textcolor{black}{#1}}

\usepackage{textcomp}

\usepackage{stmaryrd}

\usepackage{comment}

\usepackage{graphicx,subfigure}

\usepackage{stfloats}
\usepackage{tabularx }

\usepackage{makecell}

\usepackage{colortbl}

\newcolumntype{P}[1]{>{\centering\arraybackslash}p{0.66cm}}
 
\usepackage{anyfontsize}
\usepackage{t1enc}

\newcommand{\T}{^{\scriptscriptstyle\rm T}}

\usepackage[strict]{changepage}
\usepackage{relsize}

\usepackage{diagbox}

\newcommand\tblue{\cellcolor{blue!25}}
\newcommand\tred{\cellcolor{red!25}}

\begin{document}

\title{ \hig{Neural Decentralized Conditions for \\Power System Stability Analysis}}

\author{Tong Han,~\IEEEmembership{Member, IEEE}, 
        Yan Xu,~\IEEEmembership{Senior Member, IEEE}, 
        Rui Zhang,~\IEEEmembership{Member, IEEE}
\thanks{The authors are with the School of Electrical and Electronic Engineering, Nanyang Technological University, Singapore 639798. \\
This work was originally submitted to \textit{IEEE Power Engineering Letters} but was not accepted, mainly because its content better suited a full paper. It has since been extended and combined with another work into a full paper, accepted for publication in \textit{IEEE Transactions on Power Systems}~\cite{han-temp}.  
}


}

\maketitle

\begin{abstract} 
    This letter proposes a novel method to establish decentralized stability conditions (DSCs) of power systems. 
    \hig{This method formulates DSCs with meticulously crafted neural networks, based on the insight into the structure and properties of the existing analytically derived DSCs. A tailored loss function is further devised to iteratively train the neural networks, ensuring validity and reducing conservatism of the DSCs. } 
    Numerical results demonstrate that the resultant neural DSCs are less conservative than the conventional analytical DSCs, and become generalizable if a relatively large-scale system is used for training.
\end{abstract}


\IEEEpeerreviewmaketitle

\section{Introduction}

Decentralized stability analysis verifies system stability by examining whether specific conditions are satisfied independently by each group of components within the system. 
It offers a scalable fashion for stability assessment and facilitates the distributed design of stabilizing controllers \cite{4-rp-8-short}. 
Most existing decentralized stability conditions (DSCs) are analytically derived based on control theory principles, including 
the passivity theory \cite{4-rp-11, 4-rp-12}, dissipativity theory \cite{4-rp-17}, small gain theorem and small phase theorem \cite{4-rp-10}, zero exclusion principle \cite{4-1312}, and Lyapunov method \cite{4-rp-13}. 
The derivations often require a deep insight into system dynamics, which can sometimes be difficult to gain. 
More importantly, these DSCs generally face challenges in reducing conservatism \cite{4-rp-8-short}. 
In this letter, we depart from the conventional paradigm and develop a learning-based method for the DSC establishment. The DSCs are formulated using meticulously crafted neural networks (NNs) and trained iteratively using a tailored loss function. This method needs no insight into system dynamics, yet produces neural DSCs that are less conservative. 
Numerical tests also show that the neural DSCs trained for large-scale systems exhibit a level of generalizability like those analytically derived.

\section{\hig{Small-Signal Model and Stability Conditions}}

Consider a power system, where $N$ buses are interconnected via lines, and each bus is connected with a component (e.g., synchronous generator, converter, load, or their combination). 
The system dynamics can be formulated as the collection of individual dynamics at all buses as  follows: 
\begin{equation}\label{eq-8-1-1}
    \! \bm{E}_{c(i)} \dot{\bm{x}_i} =\!  f_{c(i)}( \bm{x}_i ; \bm{\rho}_i ) +\!\! \medmath{\sum_{ j \in \mathcal{A}_i }} g_{ c(i) c(j)  } ( \bm{x}_i, \bm{x}_j ; \bm{\rho}_{ij} ) ~~ \forall i \!\in\! \mathcal{N},
\end{equation} 
where $\mathcal{N} = \{1, 2, ..., N \}$ is the set of all buses; 
$c(i)$ denotes the type of component connected at bus $i$ (components with identical dynamic models are seen as the same type), and the bus connecting with a component of type $c(i)$ is referred to as a \textit{$c(i)$-type bus}; 
$\bm{x}_i \in \mathbb{R}^{n_{c(i)}}$ ($n_{c(i)}$ is the dimension of $\bm{x}_i$) is the vector collecting both state and algebraic variables associated with bus $i$ and its connected component; 
$\bm{\rho}_i \in \mathbb{R}^{m_{c(i)}}$ ($m_{c(i)}$ is the dimension of $\bm{\rho}_i$) is the vector collecting parameters of the component connected at bus $i$; 
$\bm{\rho}_{ij} \in \mathbb{R}^{m}$ ($m$ is the dimension of $\bm{\rho}_{ij}$) is the vector collecting parameters of the line between bus $i$ and $j$; 
$\bm{E}_{c(i)}$ is a square (0,1)-matrix of order $n_{c(i)}$, encapsulating the algebraic and differential structure of the dynamics at a $c(i)$-type bus; 
$\mathcal{A}_i \subset \mathcal{N}$ is the set of all buses connecting with bus $i$; 
$f_{c(i)}: \mathbb{R}^{n_{c(i)}} \mapsto \mathbb{R}^{n_{c(i)}}$ represents the local dynamics at a $c(i)$-type bus; 
$g_{ c(i) c(j)}: \mathbb{R}^{n_{c(i)}} \times \mathbb{R}^{n_{c(j)}} \mapsto \mathbb{R}^{n_{c(i)}}$ represents the coupling dynamics between a $c(i)$-type bus and a $c(j)$-type bus.  
Next, by linearizing system (\ref{eq-8-1-1}) around its equilibrium point, 
the small-signal model of the individual dynamics at any bus $i \in \mathcal{N}$ can be formulated as follows: 
\begin{equation}\label{eq-8-1-2}
    \begin{aligned}
        & \bm{E}_{c(i)} \Delta \dot{\bm{x}_i}  = \\[-2.7mm]
        & \bm{A}_{c(i)}( \bm{x}_i^*,  \bm{\rho}_i ) \!\cdot\! \Delta  \bm{x}_i \!+\! \medmath{\sum\nolimits_{j \in \mathcal{A}_i}} \!\! \bm{A}_{c(i) c(j)} ( \bm{x}_i^*, \bm{x}_j^*, \bm{\rho}_{ij}) \! \begin{bmatrix}
            \Delta  \bm{x}_i \\[-1mm] \Delta \bm{x}_j
        \end{bmatrix}
    \end{aligned}
\end{equation}
where $\bm{x}_i^*$ is the value of $\bm{x}_i$ at the equilibrium point of (\ref{eq-8-1-1}); 
$\Delta \bm{x}_i$ is the perturbation of $\bm{x}_i$ from $\bm{x}_i^*$; 
$\bm{A}_{c(i)} \!\!=\!\! \frac{ \partial f_{c(i)}( \bm{x}_i ; \bm{\rho}_i )   }{ \partial \bm{x}_i   }|_{\bm{x}_i = \bm{x}_i^*}$ and 
$\bm{A}_{c(i) c(j)} \!\!=\!\! \frac{  \partial g_{ c(i) c(j)  } ( \bm{x}_i, \bm{x}_j ; \bm{\rho}_{ij} ) }{ \partial [ \bm{x}_i\T, \bm{x}_i\T]\T  } |_{ \bm{x}_i = \bm{x}_i^*, \bm{x}_j = \bm{x}_j^* }$ are both matrix functions.  
Combining the individual small-signal model of all buses gives the small-signal model of the system, written as:  
\begin{equation}\label{eq-8-1-3}
    \bm{E} \Delta \dot{\bm{x}} = \bm{A}(\bm{x}^*, \bm{\rho}) \cdot \Delta \bm{x}
\end{equation}
where $\Delta \bm{x}$ and $\bm{x}^*$ are formed by concatenating $\Delta \bm{x}_i$ and $\bm{x}_i^*$ of all buses, respectively; 
$\bm{\rho}$ is formed by concatenating $\bm{\rho}_i$ of all buses and $\bm{\rho}_{ij}$ of all lines; 
the matrix $\bm{E}$ is formed with $\bm{E}_{c(i)}$ of all buses, and $\bm{A}$ is formed with $\bm{A}_{c(i)}$ of all buses and $\bm{A}_{c(i), c(j)}$ of all pairs of connected buses.

The sufficient and necessary conditions for asymptotic stability of system (\ref{eq-8-1-3}), i.e., small-signal stability of system (\ref{eq-8-1-1}) at the equilibrium point $\bm{x}^*$, can be expressed in a \textit{centralized} fashion. For instance, using generalized eigenvalues, it can be stated as: system (\ref{eq-8-1-3}) is asymptotically stable iff all generalized eigenvalues of the pair $(\bm{E}, \bm{A}(\bm{x}^*, \bm{\rho}) )$ exhibit a negative real part, i.e, $\lambda_{\max}(\bm{E}, \bm{A}(\bm{x}^*, \bm{\rho})) < 0$ with $\lambda_{\max}(\cdot)$ being the largest real part of the generalized eigenvalues of the pair \cite{4-rp-14}. 

The asymptotic stability conditions of system (\ref{eq-8-1-3}) can also be expressed in various \textit{decentralized} ways by utilizing different forms of local information of system dynamics. 
For example, the DSCs in \cite{4-rp-11} and \cite{4-rp-12} are defined for each bus, in terms of the local dynamics of the bus (i.e., $\bm{A}_{c(i)}$) and coupling dynamics of its connected lines (i.e., $\bm{A}_{c(i)c(j)}$ for all $j \in \mathcal{A}_j$). 
Typically, these DSCs are sufficient but not necessary for asymptotic stability. 
The most critical performance of them is conservatism. 
Our focus lies on the conservatism associated with utilizing a specific form of local information, rather than exploring less conservative forms of local information, which is beyond the scope of this work.

\section{\hig{Construction of Neural DSCs}}

\subsection{\hig{Designing of Neural DSCs}}

This study employs the same form of local system dynamics information as utilized in \cite{4-rp-11} and \cite{4-rp-12} to present the construction of neural DSCs. 
However, neural DSCs can be obtained in a similar manner when employing other forms of local information of system dynamics. 
Specifically, the goal is to find DSCs such that, 
if for each bus $i \!\in\! \mathcal{N}$, $\bm{E}_{c(i)}$, $\bm{A}_{c(i)}$, $\bm{A}_{c(j)}$, and $\bm{A}_{c(i)c(j)}$ for all $j \!\in\! \mathcal{A}_j$ together satisfy the local conditions, then the system (\ref{eq-8-1-3}) is asymptotically stable. 

We first category all lines according to the types of buses they connect. For a line connecting bus $i$ and bus $j$, we represent it as two directed lines, respectively denoted as lines $(i, j)$ and $(j, i)$. The directed lines share the same type if the types of their starting buses and ending buses match pairwise. 
It is assumed that there are a total of $L$ different types of directed lines in the system, with $\mathcal{L} = \{1,2,...,L\}$ denoting the set of all directed line types. 
Let the function $e(i, j)$ return the type of any directed line $(i,j)$. 
Also assume that there are a total of $K$ different types of buses in the system, with $\mathcal{K} = \{1,2,...,K\}$ denoting the set of all bus types. 

For the same type of directed lines, their associated matrix functions $\bm{A}_{c(i)c(j)}$ and $\bm{A}_{c(i)}$ are respectively the same matrix functions with different input values. Thus we first introduce an NN for each type of directed lines to map the inputs of $\bm{A}_{c(i)c(j)}$ and $\bm{A}_{c(i)}$ (i.e., $\bm{x}_i^*$, $\bm{x}_j^*$, $\bm{\rho}_{i}$, and $\bm{\rho}_{ij}$) of each directed line into the same representation space, formulated as 
\begin{equation}
    \bm{r}_{ij} = \phi_{l}( \bm{y}_{ij} ; \bm{\zeta}_l) ~~ \forall l \in \mathcal{L}
\end{equation}
where $\bm{y}_{ij} = [{\bm{x}_i^*}\T, {\bm{x}_j^*}\T, \bm{\rho}_{i}\T, \bm{\rho}_{ij}\T ]\T \in \mathbb{R}^{\eta_l}$, with $(i, j)$ be any directed line of type $l$ and $\eta_l = n_{c(i)} + n_{c(j)} + m_{c(i)} + m$; 
$\bm{r}_{ij} \in \mathbb{R}^{\eta_{\rm r}}$ is the $\eta_{\rm r}$-dimensional representation vector for the directed line $(i, j)$; 
$\phi_{l}: \mathbb{R}^{\eta_l}  \to \mathbb{R}^{\eta_{\rm r}} $, represents an NN with $\bm{\zeta}_l$ being its combined parameter vector. 

Similarly, we further introduce an NN for each bus type to map the inputs of $\bm{A}_{i}$ (i.e, $\bm{x}_i^*$ and $\bm{\rho}_i$) of the bus, together with the aggregation of the representation vectors of all directed lines starting from the bus, into a scalar to indicate satisfaction of stability conditions at the bus. 
Formally, for each bus type $k \in \mathcal{K}$, its stability condition candidate is represented as 
\begin{equation}
    \Phi_{k}( \bm{y}_{i} ; \bm{\theta}_k )  < 0 
\end{equation} 
where 
$\bm{y}_{i} \!=\! [{\bm{x}_i^*}\T, \bm{\rho}_{i}\T, \bm{a}_{i}\T ]\T \!\in\! \mathbb{R}^{\mu_k}$, with $i$ being any bus of type $k$, $\mu_k \!=\! n_{c(i)}  \!+ m_{c(i)} \!+ 3 \eta_{\rm r}$, and $\bm{a}_i \!=\! [\sum_{j \in \mathcal{A}_i} \! \bm{r}_{ij}\T, \frac{1}{|\mathcal{A}_i|} \sum_{j \in \mathcal{A}_i} \!\bm{r}_{ij}\T,$ $ \max\nolimits_{j \in \mathcal{A}_i} \bm{r}_{ij}\T  ]\T $; 
$\Phi_{k}:\! \mathbb{R}^{\mu_k}  \!\to\! \mathbb{R} $, is an NN with the combined parameter vector $\bm{\theta}_k$. 

\subsection{\hig{Training of Neural DSCs}}

\hig{With the above design}, establishing the DSCs amounts to searching values of $\bm{\zeta}_l$ and $\bm{\theta}_k$, such that
\begin{equation}\label{eq-8-1-5}
         \{ \forall i \!\in\! \mathcal{N}, \Phi_{c(i)}( \bm{y}_{i} ; \bm{\theta}_{c(i)} )  \!<\! 0\}  \!\Rightarrow\! \lambda_{\max}(\bm{E}, \bm{A}(\bm{x}^*, \bm{\rho})) \!<\! 0
\end{equation} 
for any $(\bm{x}^*, \bm{\rho}) \in \mathbb{X} \times \mathbb{P}$, with $\mathbb{X}$ and $\mathbb{P}$ being the regions of $\bm{x}^*$ and $\bm{\rho}$ of interest respectively. 
Further, we divide the region $\mathbb{X} \times \mathbb{P}$ into two parts: the unstable region $\mathbb{Y}_1$, where system (\ref{eq-8-1-3}) is not asymptotically stable; and the stable region $\mathbb{Y}_2$, where system (\ref{eq-8-1-3}) is asymptotically stable. 
Then, proposition (\ref{eq-8-1-5}) yields the following corollaries: 
\begin{itemize}

    \item The probability that the left-hand side (LHS) of (\ref{eq-8-1-5}) holds for any randomly chosen $(\bm{x}^*, \bm{\rho}) \in \mathbb{Y}_1$, denoted as $P_1$, is 0. This is the equivalent contrapositive of proposition (\ref{eq-8-1-5}) and should therefore be guaranteed strictly.

    \item The probability that the LHS of (\ref{eq-8-1-5}) does not hold for any $(\bm{x}^*, \bm{\rho}) \!\!\in\!\! \mathbb{Y}_2$, denoted as $P_2$, is indeterminate. This probability quantifies the DSCs' conservatism, with smaller values indicating lower conservatism overall.

\end{itemize}

\begin{figure}[t!]
	\centering 
    \includegraphics[scale=0.8]{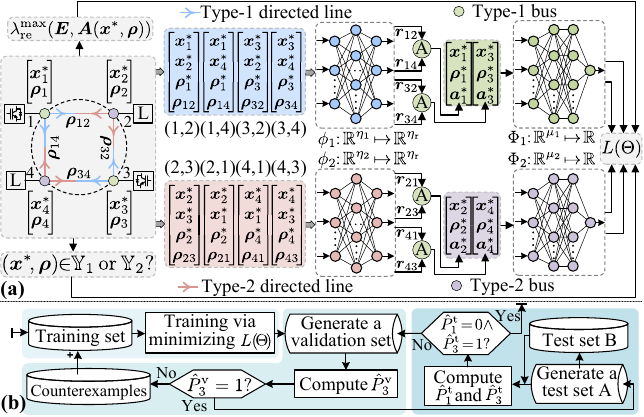}   
	\caption{(a) Illustration of the designed NNs and the loss function; (b) flowchart of iterative training of the NNs.}
	\label{fig-8-1}
\end{figure}

Next, we collect $N_1 + N_2$ samples of $(\bm{x}^*, \bm{\rho})$ to form the training set, with the first $N_1$ samples in region $\mathbb{Y}_1$ and the remaining $N_2$ samples in $\mathbb{Y}_2$.  
According to the above corollaries, 
the values of $\bm{\zeta}_l$ and $\bm{\theta}_k$ with which (\ref{eq-8-1-5}) holds and the DSCs are minimally conservative for all training samples, are given by the following optimization problem: 
\begin{equation}\label{eq-8-1-7}
    \begin{aligned}
        \min\nolimits_{\Theta} ~ & \hat{P}_2 =   \tfrac{1}{ N_2 } \medmath{\sum\nolimits_{ s = N_1+1 }^{N_1 + N_2}}  \max_{i \in \mathcal{N}}   H\big(  \Phi_{c(i)}( \bm{y}_{i}^s ; \bm{\theta}_{c(i)} ) \big)   \\[-1mm]
        \text{s.t.} ~  & \hat{P}_1 \!=\! \tfrac{1}{ N_1 } \! \medmath{\sum\nolimits_{ s = 1 }^{N_1}}  \Big[ 1 \!-\! \max_{i \in \mathcal{N}} H \big( \Phi_{c(i)}( \bm{y}_{i}^s ; \bm{\theta}_{c(i)} ) \big)  \Big] \!=\! 0
    \end{aligned}
\end{equation}
where $\Theta = [\bm{\zeta}_1\T, \bm{\zeta}_2\T, \cdots, \bm{\zeta}_L\T, \bm{\theta}_1\T, \bm{\theta}_2\T, \cdots, \bm{\theta}_K\T ]\T$, 
$H(\cdot)$ is the unit step function that equals to $0$ for negative arguments and 1 otherwise; 
$\bm{y}_{i}^{s}$, $\bm{x}^{*, s}$, and $\bm{\rho}^s$ denote separately the values of $\bm{y}_{i}$, $\bm{x}^{*}$, and $\bm{\rho}$ of the $s$-th sample. 
The terms $\hat{P}_1$ and $\hat{P}_2$ are the empirical values of probability $P_1$ and $P_2$, respectively. 

To solve problem (\ref{eq-8-1-7}), all NNs are trained together using the loss function designed as follows:
\begin{equation}
    L(\Theta) = w_1 L_1 + w_2 L_2 + w_{\rm aux} L_{\rm aux} ~  \text{with}
\end{equation}
\begin{equation}
    L_1 = - \tfrac{1}{ N_1 } \medmath{\sum\nolimits_{ s = 1 }^{N_1}} \log \Big[ \max_{i \in \mathcal{N}} \sigma \big( \Phi_{c(i)}( \bm{y}_{i}^s ; \bm{\theta}_{c(i)} ) \big)  \Big]
\end{equation}
\begin{equation}
    L_2 = - \tfrac{1}{ N_2 }  \medmath{\sum\nolimits_{ s = N_1+1 }^{N_1 + N_2}} \log \Big[ 1 - \max_{i \in \mathcal{N}} \sigma \big( \Phi_{c(i)}( \bm{y}_{i}^s ; \bm{\theta}_{c(i)} ) \big)  \Big]
\end{equation}
\begin{equation} 
    \begin{aligned}
        & L_{\rm aux}  =  1/(N_1 + N_2) \cdot   \medmath{\sum\nolimits_{ s = 1 }^{N_1 + N_2}}  \big[ \\[-1.5mm] 
        &  \max\nolimits_{i \in \mathcal{N}}  \Phi_{c(i)}( \bm{y}_{i}^s ; \bm{\theta}_{c(i)} ) - \lambda_{\max}(\bm{E}, \bm{A}(\bm{x}^{*, s}, \bm{\rho}^s ))  \big]^2   
    \end{aligned} 
\end{equation}
where $w_1$, $w_2$, and $w_{\rm aux}$ are scalar weights; $\sigma(\cdot)$ is the sigmoid function; 
the terms $L_1$ and $L_2$, devised with inspiration from the cross-entropy loss, each serve as proxies for $\hat{P_1}$ and $\hat{P_2}$. 
The term $L_{\rm aux}$ represents the average distance between the largest value of $\Phi_{c(i)}( \cdot )$ among all buses (roughly correlating with the system's stability level), and $\lambda_{\max}(\cdot)$, over all samples. 
This term offers some physics-informed significance to the conditions, thereby providing additional training guidance. 
Moreover, $\omega_1$ is updated every few training epochs using the Lagrangian duality method \cite{4-1688}; $\omega_{\rm aux}$ is updated using an exponential decay schedule; 
and $\omega_2$ is updated using an exponential growth schedule, with updates occurring each time $\hat{P_1} = 0$ and the epoch number is less than a preset value.

Fig. \ref{fig-8-1}-(a) using a 4-bus system illustrates the designed NNs and the loss function. 
This system comprises two types of buses and two types of directed lines: 
type-1 buses including 1 and 3, and type-2 buses including 2 and 4, 
type-1 directed lines including (1,2), (1,4), (3,2), and (3,4), and type-2 directed lines including (2,3), (2,1), (4,1), and (4,3). 
Thus, two NNs, i.e., $\phi_1$ and $\phi_2$ are introduced for mapping line information, and two NNs, i.e., $\Phi_1$ and $\Phi_2$ are introduced as the stability condition candidates respectively for type-1 buses and type-2 buses. 
For each sample of $(\bm{x}^*, \bm{\rho})$, taking bus 1 for example,  $\bm{y}_{12}$ and $\bm{y}_{14}$, via $\phi_1$, are separately mapped as representation vectors $\bm{r}_{12}$ and $\bm{r}_{14}$, which are further aggregated as $\bm{a}_1$; and then $\bm{a}_1$ together with the local dynamic information of bus 1 serves as the input of $\Phi_1$, which outputs a stability indicator. Analogous outputs for all buses, the value of $\lambda_{\max}$, and stability for the sample, are together used for forming the loss function $L(\Theta)$.

To ensure proposition (\ref{eq-8-1-5}) holds for any $(\bm{x}^*, \bm{\rho}) \in \mathbb{X} \times \mathbb{P}$ almost surely, the NN training iterates with new samples from validation sets added into the training set, as shown in Fig. \ref*{fig-8-1}-(b). 
Specifically, with fixed NNs $\Phi_k$ for all $k \in \mathcal{K}$ after training, the probability that proposition (\ref{eq-8-1-5}) holds for any $(\bm{x}^*, \bm{\rho}) \in \mathbb{X} \times \mathbb{P}$, denoted as $P_3$, can be estimated. 
Generating $N_{\rm v}$ samples of $(\bm{x}^*, \bm{\rho})$ where the LHS of (\ref{eq-8-1-5}) holds to form a validation set, let $N_{\rm c}$ denote the number of counterexamples, i.e., validation samples where $\lambda_{\max}(\bm{E}, \bm{A}(\bm{x}^{*}, \bm{\rho} )) \!\geq\! 0$. 
Then the empirical values of $P_3$ on the validation set is expressed as $\hat{P}_3^{\rm v} \!\!=\!\! (N_{\rm v} \!-\! N_{\rm c})/{N_{\rm v}}$. 
If $\hat{P}_3^{\rm v} \!<\! 1$, the counterexamples found together with some samples around them are added to the training set to retrain the NNs; otherwise, i.e., no counterexamples exist, the correctness of the obtained DSCs is further tested using two test sets, named A and B. 
Set A is analogous to the validation set but significantly larger in size. The empirical value of $P_3$ on test set A, denoted as $\hat{P}_{3}^{\rm t}$, is analogous to $\hat{P}_3^{\rm v}$. 
Test set B contains all counterexamples found in test set A. An empirical value of $P_1$, denoted as $\hat{P}_1^{\rm t}$, can be easily obtained on this test set. If $\hat{P}_1^{\rm t} = 0$ and $\hat{P}_3^{\rm t} = 1$, the obtained DSCs are almost certainly correct; otherwise, the validation set is regenerated to find counterexamples for retraining the NNs. 
It is noted that in each iteration, test set A needs to be regenerated since the NNs are changed. 
Test set B combines all counterexamples found in test set A from both the current and all previous iterations.

\section{Case Study}

\begin{figure}[t!]
	\centering 
    \includegraphics[scale=0.83]{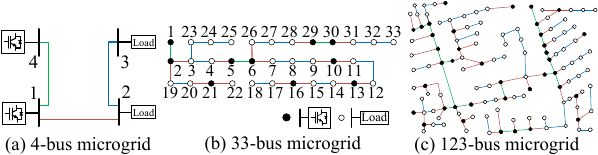}  
	\caption{Diagrams of the three test systems.}
	\label{fig-8-1-5}
\end{figure}

The proposed method is tested using the three microgrid systems shown in Fig. \ref{fig-8-1-5}. 
Each system contains two types of buses: buses connected with an inverter with $p$-$\omega$/$q$-$v$ droop control, and buses connected with an $\omega$- and $v$-dependent load. The region of interest for the same type of parameters is set to be identical. The activation functions of all NNs are set as ReLU. 
Adam with step decay of learning rate is used for training. 
The sizes of training, validation, and test (A) set are separately 20k, 20k, and 40k. 
The batch size is 2560.

\begin{figure}[t!]
	\centering 
    \includegraphics[scale=1.17]{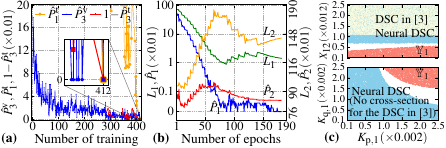}  
	\caption{(a)-(b) Curves of the associated terms during the training; (c) cross-sections of the stability regions formed by different DSCs for the 4-bus system.}
	\label{fig-8-1-6}
\end{figure}

Fig. \ref{fig-8-1-6}-(a) and \ref{fig-8-1-6}-(b), taking the 123-bus system for example, respectively show the curves of $\hat{P}_3^{\rm v}$, $\hat{P}_1^{\rm t}$, and $\hat{P}_3^{\rm t}$ during the entire iterative training process, and the curves of $L_1$, $L_2$, $\hat{P}_1$, and $\hat{P}_2$ for the last time of training. 
It can be seen that for the last time of training, $\hat{P}_1$ is forced to 0 and $\hat{P}_2$ converges to a value less than 1 by training the NNs using the devised loss function, indicating problem (\ref{eq-8-1-7}) is solved effectively. 
Also, $\hat{P}_1^{\rm t}$ and $\hat{P}_3^{\rm t}$ simultaneously reach 0 and 1, and thus the iterative training scheme yields correct DSCs for the test system. 
Furthermore, 
Table \ref{tab-5-8-1} shows the performance of both the neural DSCs and the passivity-based DSC in \cite{4-rp-12}. This evaluation is based on the empirical values of ${P}_1$, estimated on the final test set B from the training for the associated test system, as well as $P_2$ and $P_3$, both computed using a 40k-sized sample set generated similarly to the training set. The neural DSCs established for each test system are evaluated on all test systems. 
In Table \ref{tab-5-8-1}, the empirical values of $P_1$ and $P_2$ are 0 and 1, respectively, for the neural DSC obtained and evaluated using the same test system. This again validates the correctness of each neural DSC with respect to the system used to train it. Interestingly, the neural DSC from the 123-bus system is also correct with respect to the other two systems, given the unchanged empirical values of $P_1$ and $P_2$ evaluating on the other two systems. 
Further, the correctness of the neural DSC from the 123-bus system still holds in nearly all cases when the test system removes one bus, according to the last row's results in Table \ref{tab-5-8-1}. 
Therefore, the neural DSC from the 123-bus system is a generalizable DSC similar to conventional DSCs rather than one only holding for a particular system. 
In addition, the neural DSCs are all less conservative than the passivity-based DSC in \cite{4-rp-12}, given the empirical values of $P_2$ in Table \ref{tab-5-8-1}. Intuitively, Fig. \ref{fig-8-1-6}-(c) displays two cross-sections of different stability regions for the 4-bus test system, demonstrating the correctness and low conservatism of the obtained neural DSCs.

\begin{table}[t!]
    \centering
 
    \caption{Performance of the neural DSCs and the DSC in \cite{4-rp-12} }
    \setlength{\tabcolsep}{0.99pt} 

    \setlength{\aboverulesep}{0.2pt}
    \setlength{\belowrulesep}{0.2pt}
    \setlength{\extrarowheight}{-.0ex}
 
    \footnotesize{
    \begin{tabular*}{\hsize}{@{}p{1.06cm}ccccccccc}\toprule
    \multirow{2}{*}{ DSCs  } & \multicolumn{3}{c}{$P_1$} & \multicolumn{3}{c}{$P_3$} & \multicolumn{3}{c}{$P_2$}  
    \\ \cmidrule(lr){2-4} \cmidrule(lr){5-7} \cmidrule(lr){8-10} 
                        & 4-bus       & 33-bus     & 123-bus    &  4-bus    & 33-bus   & 123-bus  & 4-bus & 33-bus & 123-bus      \\ \midrule[0.7pt]
    4-bus               &  \tblue0.00\% & 40.4\%       & 67.4\%       &  \tred100\% & 63.1\%     & 6.98\%     &  95.5\% & ---    & ---   \\
    33-bus              &  17.2\%       & \tblue0.00\% & 31.7\%       &  97.3\%     & \tred100\% & 23.6\%     &  ---  & 46.3\%    & ---   \\
    123-bus             &  \tblue0.00\% & \tblue0.00\% & \tblue0.00\% &  \tred100\% & \tred100\% & \tred100\% &  91.6\% & 38.2\%    & 15.9\%   \\
    \cite{4-rp-12}      &  \tblue0.00\% & \tblue0.00\% & \tblue0.00\% &  \tred100\% & \tred100\% & \tred100\% &  82.1\% & 25.6\%    & 6.55\%   \\ 
    123-bus$^*$         &  \tblue 2/2 & \tblue5/5 & \tblue40/42 &  \tblue 2/2 & \tblue5/5 & \tblue40/42 &  93.2\% & 38.5\%     & 16.1\%   \\ \bottomrule
    \multicolumn{10}{@{}p{1\columnwidth}@{}}{
        \footnotesize{\textit{Note}: The last row shows the proportion of $P_1 = 0$ for $P_1$, the proportion of $P_3 = 100\%$ for $P_3$, and the mean value for $P_2$, among all cases when the associated test system remove one bus without disconnecting the network. 
      }}
    \end{tabular*} 
    }
    \label{tab-5-8-1}  
\end{table}

\section{Conclusion}

This letter offers a novel learning-based paradigm for the DSC establishment of power systems. 
The neural DSCs are less conservative compared to those derived analytically, while they can still be generalizable. 
Future works will explore the requirement for the system used for training and methodological improvement to obtain globally generalizable neural DSCs, and neural DSCs for more complex system dynamics.

\ifCLASSOPTIONcaptionsoff
  \newpage
\fi

\vspace{-6pt}
\bibliographystyle{IEEEtran}
\bibliography{4.bib}

\end{document}